*Research Article*

# The Accurate Modification of Tunneling Radiation of Fermions with Arbitrary Spin in Kerr-de Sitter Black Hole Space-Time

**Bei Sha , Zhi-E Liu, Xia Tan, Yu-Zhen Liu, and Jie Zhang**

*School of Physics and Electronic Engineering, Qilu Normal University, Jinan 250200, China*

Correspondence should be addressed to Bei Sha; shabei@qlnu.edu.cn

The quantum tunneling radiation of fermions with arbitrary spin at the event horizon of Kerr-de Sitter black hole is accurately modified by using the dispersion relation proposed in the study of string theory and quantum gravitational theory. The derived tunneling rate and temperature at the black hole horizons are analyzed and studied.

## 1. Introduction

Lorentz dispersion relation has been regarded as the basic relation of modern physics, which is related to general relativity and quantum field theory. However, the development of quantum gravity theory shows that Lorentz relation must be modified in the high energy field. The dispersion relation theory in the high-energy field has not been established and needs to be further studied. Nevertheless, it is generally believed that the magnitude of such correction is only the Planck scale [1–10]. The dynamical equation of the spin-1/2 fermion is the Dirac equation in curved space-time, while that of spin-3/2 fermion is described by the Rarita–Schwinger equation in curved space-time. In 2007, Kerner and Mann et al. proposed to study the quantum tunneling of spin-1/2 fermion using semiclassical theories and methods [11, 12]. Subsequently, Yang and Lin studied the quantum tunneling of fermions and bosons applying the semiclassical Hamilton–Jacobi method, and found that the behaviours of fermions and bosons could both be described by one same equation—Hamilton–Jacobi equation, and studied the quantum tunneling radiation of various black holes with the Hamilton–Jacobi theory and method [13, 14]. In the process of studying the thermodynamics of black holes, it is worth mentioning that Banerjee and Majhi et al. put forward Hamilton–Jacobi method beyond the semiclassical approximation to modify the quantum tunnelling of bosons and fermions, and then studied the temperature, entropy, and other physical quantities of black holes [15, 16]. Zhao et al. conducted effective research on Hawking radiation of various black holes [17–19]. What is worth studying is that Parikh and Wilczek corrected the particle tunneling rate at the event horizon of the black hole by taking into account the special and real situation that the background space-time changes before and after particle tunneling [20]. Banerjee and Majhi et al. Further developed the mechanism of quantum tunneling through chirality method and semiclassical approximation [21–25]. Other scholars have also conducted a series of effective studies on the quantum tunneling rate and entropy of various black holes [26–43]. Science and technology are always developing and progressing with the passage of time. The research on theoretical physics, astrophysics and related hot topics can always promote the continuous improvement of scientific research and bring about knowledge innovation. Recent studies show that the modification of Lorentz dispersion relation is necessary to modify the particle dynamical equation in the space-time of strong gravitational field, which is mainly to correct the quantum tunneling of fermions or bosons at the event horizon of the black hole, as well as the gravitational wave equation in curved space-time can also be modified as necessary. The purpose of this paper is to analyze the physical quantities, such as quantum tunneling rate of arbitrary spin fermions, black hole temperature, and black hole entropy in Kerr-de Sitter black hole space-time. Therefore, in Section 2, the Rarita–Schwinger equation, which is the dynamical equation of arbitrary spin fermions, will be modified by applying the modified Lorenz dispersion relation. In Section 3 the quantum tunneling radiation of arbitrary spin fermions at horizons of Kerr-de Sitter black hole, which is a rotating, stationary, axisymmetric, and with a cosmic horizon black hole, is studied. Section 4 includes some conclusions and discussions.

## 2. From the Rarita–Schwinger Equation in Flat Space-Time to the Rarita–Schwinger Equation in Curved Space-Time

A modified dispersion relation on the quantum scale is given by [1–10]

$$p_0^2 = p^2 + m^2 - (Lp_0)^\alpha p^2, \tag{1}$$

Consider the special case of the modified dispersion relation (the speed of light in vacuum $c$ equals unit in the notations), where $p_0$ and $p$ are the energy and momentum of particle and $L$ is a constant on the Planck scale. When considering $\alpha = 2$, the dynamical equation of the spin-1/2 fermion is known as the Dirac equation. In flat space-time it is expressed as

$$\left(\bar{\gamma}^\mu \partial_\mu + \frac{m}{\hbar} - iL\bar{\gamma}^t \partial_t \bar{\gamma}^j \partial_j\right)\Psi = 0. \tag{2}$$

In flat space-time, the more general fermion dynamical equation was expressed by Rarita–Schwinger as [44]

$$\left(\bar{\gamma}^\mu \partial_\mu + \frac{m}{\hbar}\right)\Psi_{\alpha_1 \cdots \alpha_k} = 0, \tag{3}$$

which satisfies the conditions of $\bar{\gamma}^\mu \Psi_{\mu\alpha_2 \cdots \alpha_k} = \partial_\mu \Psi^\mu_{\alpha_2 \cdots \alpha_k} = \Psi^\mu_{\mu\alpha_3 \cdots \alpha_k} = 0$. In Rarita–Schwinger equation in flat space-time, generalizing the ordinary differential into the covariant differential, and the ordinary derivative into the covariant derivative, the Rarita–Schwinger equation of arbitrary spin fermions in curved space-time can be obtained as

$$\left(\gamma^\mu D_\mu + \frac{m}{\hbar}\right)\Psi_{\alpha_1 \cdots \alpha_k} = 0, \tag{4}$$

and it meets the condition of

$$\gamma^\mu \Psi_{\mu\alpha_2 \cdots \alpha_k} = D_\mu \Psi^\mu_{\alpha_2 \cdots \alpha_k} = \Psi^\mu_{\mu\alpha_3 \cdots \alpha_k} = 0, \tag{5}$$

where the Gamma matrix satisfies the following commutation relation

$$\gamma^\mu \gamma^\nu + \gamma^\nu \gamma^\mu = 2g^{\mu\nu} I, \tag{6}$$

and $D_\mu$ is the operator of covariant derivative in curved space-time, i.e

$$D_\mu = \partial_\mu + \Omega_\mu + \frac{i}{\hbar} e A_\mu, \tag{7}$$

where $\Omega_\mu$ is spin connection in curved space-time.

According to formula (1), taking $\alpha = 2$, we can rewrite the Rarita–Schwinger equation in flat space-time as

$$\left(\bar{\gamma}^\mu \partial_\mu + \frac{m}{\hbar} - \sigma\hbar\bar{\gamma}^t \partial_t \bar{\gamma}^j \partial_j\right)\Psi_{\alpha_1 \cdots \alpha_k} = 0. \tag{8}$$

For the correction on quantum scale, take $\sigma$ as a small correction term $\sigma \ll 1$, therefore, $\sigma\hbar\bar{\gamma}^t \partial_t \bar{\gamma}^j \partial_j \Psi_{\alpha_1 \cdots \alpha_k}$ is very small. The matrix Equation (8) is generalized to Kerr-de Sitter curved space-time, obtaining

$$\left(\gamma^\mu D_\mu + \frac{m}{\hbar} - \sigma\hbar\gamma^t D_t \gamma^j D_j\right)\Psi_{\alpha_1 \cdots \alpha_k} = 0. \tag{9}$$

This matrix equation can only be solved in a specific curved space-time. For this purpose, the Fermion wave function is set as

$$\Psi_{\alpha_1 \cdots \alpha_k} = \xi_{\alpha_1 \cdots \alpha_k} e^{(i/\hbar)S}, \tag{10}$$

where $S$ is the action of fermion with mass $m$. The line element of Kerr-de Sitter black hole in the Boyer–Lindquist coordinates is

$$ds^2 = -\frac{\Delta_r}{\chi^2 \rho^2}\left(dt - a\sin^2\theta d\varphi\right)^2 + \frac{\Delta_\theta \sin^2\theta}{\chi^2 \rho^2}\left[adt - (r^2 + a^2)d\varphi\right]^2 + \rho^2\left(\frac{dr^2}{\Delta_r} + \frac{d\theta^2}{\Delta_\theta}\right), \tag{11}$$

which describes a rotating black hole with a cosmic horizon, and where

$$\rho^2 = r^2 + a^2\cos^2\theta,$$
$$\Delta_r = (r^2 + a^2)\left(1 - \frac{r^2}{l^2}\right) - 2Mr,$$
$$\Delta_\theta = 1 + \frac{a^2}{l^2}\cos^2\theta,$$
$$\chi = 1 - \frac{a^2}{l^2}, \tag{12}$$

where $l$ is the radius of curvature of the cosmic horizon. Obviously, the event horizon $r_H$ and the cosmic horizon $r_c$ of this black hole satisfy the equation, respectively, as

$$\Delta_r(r_H) = 0, \ \Delta_r(r_c) = 0. \tag{13}$$

From (11), (12), and (13), we can get

$$\partial_\varphi S = j, \ \partial_t S = -\omega, \tag{14}$$

where $\omega$ is the energy and $n$ is a component of the generalized angular momentum of the particles tunneling from the black hole. The electromagnetic potential of the particles $A_\mu = 0$ in this space-time can also be known from expression (11). To solve the matrix Equation (9), suppose

$$\Gamma^\mu = i\gamma^\mu - \sigma\omega\gamma^t\gamma^\mu,$$
$$m_k = m - \sigma g^{tt}\omega. \tag{15}$$

Substituting formulas (14) and (15) into Equation (9), and Equation (9) is simplified as

$$i\gamma^\mu \partial_\mu S\xi_{\alpha_1 \cdots \alpha_k} + m_k \xi_{\alpha_1 \cdots \alpha_k} - \sigma\gamma^t \omega \gamma^j \partial_j S\xi_{\alpha_1 \cdots \alpha_k}$$
$$= \Gamma^\mu \partial_\mu S\xi_{\alpha_1 \cdots \alpha_k} + m_k \xi_{\alpha_1 \cdots \alpha_k} = 0. \tag{16}$$

From Equation (16) we can obtain

$$\Gamma^\nu \Gamma^\mu \partial_\nu S\partial_\mu S\xi_{\alpha_1 \cdots \alpha_k} - m_k^2 \xi_{\alpha_1 \cdots \alpha_k} = 0, \tag{17}$$

$$\Gamma^\mu \Gamma^\nu \partial_\mu S\partial_\nu S\xi_{\alpha_1 \cdots \alpha_k} - m_k^2 \xi_{\alpha_1 \cdots \alpha_k} = 0. \tag{18}$$

From Equations (17), (18), and (6) we can get

$$[g^{\mu\nu}\partial_\mu S\partial_\nu S + m^2 - 2m\sigma g^{tt}\omega + 2i\sigma\omega g^{t\beta}\partial_\beta S\gamma^\mu\partial_\mu S \\ + \sigma^2\omega^2(g^{tt})^2 - \sigma^2\omega^2(g^{t\beta})^2(\partial_\beta S)^2]\xi_{\alpha_1\cdots\alpha_k} = 0. \quad (19)$$

This matrix equation can be further simplified as

$$i\sigma\gamma^\mu\partial_\mu S\xi_{\alpha_1\cdots\alpha_k} + m_d\xi_{\alpha_1\cdots\alpha_k} = 0. \quad (20)$$

where

$$m_d = \frac{g^{\mu\nu}\partial_\mu S\partial_\nu S + m^2 - \omega\tilde{\sigma}}{2\omega g^{t\beta}\partial_\beta S},$$

$$\tilde{\sigma} = 2m\sigma g^{tt} - \sigma^2(g^{tt})^2\omega + \sigma^2\omega(g^{t\beta}\partial_\beta S)^2. \quad (21)$$

Multiplying both sides of Equation (20) by $-i\sigma\gamma^\nu\partial_\nu S$, then

$$[\sigma^2\gamma^\nu\gamma^\mu\partial_\nu S\partial_\mu S + m_d^2]\xi_{\alpha_1\cdots\alpha_k} = 0. \quad (22)$$

The equivalent equation is

$$[\sigma^2\gamma^\mu\gamma^\nu\partial_\mu S\partial_\nu S + m_d^2]\xi_{\alpha_1\cdots\alpha_k} = 0. \quad (23)$$

From Equations (21), (22), and (23) above, it can be obtained that

$$\{[g^{\mu\nu}\partial_\mu S\partial_\nu S + m^2 - \omega\tilde{\sigma}]^2 - \sigma^2 m^2(2\omega g^{t\beta}\partial_\beta S)^2\}\xi_{\alpha_1\cdots\alpha_k} = 0. \quad (24)$$

The matrix Equation (24) has a nontrivial solution if the value of the determinant corresponding to the eigenmatrix in the equation is zero, i.e.

$$g^{\mu\nu}\partial_\mu S\partial_\nu S + m^2 - 2\sigma m\omega[g^{tt}(1+\omega) - jg^{t\varphi}] \\ - \sigma^2\omega^2[\omega^2(g^{tt})^2 - (g^{tt})^2 + (jg^{t\varphi})^2] = 0. \quad (25)$$

This equation is actually the dynamical equation of arbitrary spin fermions in the black hole space-time represented by (11) and (12). This is a precisely modified particle dynamical equation. When the correction term is ignored, the equation reverts to the Hamilton–Jacobi equation for particles of mass m in curved space-time expressed in Equations (11) and (12). Therefore, we can think of Equation (25) as the modified Rarita-Schwinger equation. The process of solving Equation (25) is the process of solving Equation (9). We only need to solve Equation (25) to find the fermion action $S$ and consequently study the characteristics of quantum tunneling radiation at the horizons of the black hole.

## 3. Tunneling Radiation Characteristics of Arbitrary Spin Fermions in Kerr-de Sitter Black Hole Space-Time

From Equations (11) and (12), the values of metric determinant and the nonzero components of the contravariant metric tensor of the space-time can be calculated, respectively, as

$$g = -\frac{1}{\chi^4}\rho^4\sin^2\theta,$$

$$g^{00} = -\frac{\chi^2}{\Delta_r\Delta_\theta\rho^2}[\Delta_\theta(r^2+a^2)^2 - \Delta_r a^2\sin^2\theta],$$

$$g^{11} = \frac{\Delta_r}{\rho^2},$$

$$g^{22} = \frac{\Delta_\theta}{\rho^2},$$

$$g^{33} = -\frac{\chi^2(a^2\Delta_\theta\sin^2\theta - \Delta_r)}{\Delta_r\Delta_\theta\rho^2\sin^2\theta},$$

$$g^{03} = g^{30} = \frac{\chi^2 a[\Delta_r - (r^2+a^2)\Delta_\theta]}{\Delta_r\Delta_\theta\rho^2}. \quad (26)$$

Substituting formula (26) into Equation (25), and Equation (25) becomes

$$-\frac{\chi^2}{\Delta_r\Delta_\theta\rho^2}[\Delta_\theta(r^2+a^2)^2 - \Delta_r a^2\sin^2\theta]\left(\frac{\partial S}{\partial t}\right)^2 \\ + 2\frac{\chi^2 a[\Delta_r - (r^2+a^2)\Delta_\theta]}{\Delta_r\Delta_\theta\rho^2}\frac{\partial S}{\partial t}\frac{\partial S}{\partial \varphi} \\ + \frac{\Delta_r}{\rho^2}\left(\frac{\partial S}{\partial r}\right)^2 + \frac{\Delta_\theta}{\rho^2}\left(\frac{\partial S}{\partial \theta}\right)^2 - \frac{\chi^2(a^2\Delta_\theta\sin^2\theta - \Delta_r)}{\Delta_r\Delta_\theta\rho^2\sin^2\theta}\left(\frac{\partial S}{\partial \varphi}\right)^2 \\ + m^2 - 2m\sigma\omega\left\{-\frac{\chi^2}{\Delta_r\Delta_\theta\rho^2}[\Delta_\theta(r^2+a^2)^2 - \Delta_r a^2\sin^2\theta]\right. \\ \left.\cdot(1+\omega) - j\frac{\chi^2 a[\Delta_r - (r^2+a^2)\Delta_\theta]}{\Delta_r\Delta_\theta\rho^2}\right\} \\ - \sigma^2\omega^2\left\{\left(-\frac{\chi^2}{\Delta_r\Delta_\theta\rho^2}\right)^2[\Delta_\theta(r^2+a^2)^2 - \Delta_r a^2\sin^2\theta]^2\right. \\ \left.\cdot(\omega^2-1) + \left[j\frac{\chi^2 a[\Delta_r - (r^2+a^2)\Delta_\theta]}{\Delta_r\Delta_\theta\rho^2}\right]^2\right\} = 0. \quad (27)$$

Multiplying both sides of Equation (27) by $\Delta_r\Delta_\theta\rho^2$, we get

$$\Delta_r^2\Delta_\theta\left(\frac{\partial S}{\partial r}\right)^2 + \Delta_r\Delta_\theta^2\left(\frac{\partial S}{\partial \theta}\right)^2 + m^2\Delta_r\Delta_\theta\rho^2 \\ - \chi^2\left\{[\Delta_\theta(r^2+a^2)^2 - \Delta_r a^2\sin^2\theta]\omega^2\right. \\ \left. -2aj\omega[\Delta_r - (r^2+a^2)\Delta_\theta] + \frac{j^2(a^2\Delta_\theta\sin^2\theta - \Delta_r)}{\sin^2\theta}\right\} \\ + 2m\sigma\omega\chi^2\{[\Delta_\theta(r^2+a^2)^2 - \Delta_r a^2\sin^2\theta] \\ \cdot(1+\omega) + ja[\Delta_r - (r^2+a^2)\Delta_\theta]\} - \sigma^2\omega^2\frac{\chi^4}{\Delta_r\Delta_\theta\rho^2} \\ \cdot\{[\Delta_\theta(r^2+a^2)^2 - \Delta_r a^2\sin^2\theta]^2(\omega^2-1) \\ + j^2 a^2[\Delta_r - (r^2+a^2)\Delta_\theta]^2\} = 0. \quad (28)$$

Dividing both sides of Equation (28) by $\Delta_\theta$, we get

$$\Delta_r^2\left(\frac{\partial S}{\partial r}\right)^2 + \Delta_r\Delta_\theta\left(\frac{\partial S}{\partial \theta}\right)^2 + m^2\Delta_r\rho^2$$
$$-\frac{\chi^2}{\Delta_\theta}\left\{\left[\Delta_\theta(r^2+a^2)^2 - \Delta_r a^2\sin^2\theta\right]\omega^2\right.$$
$$-2aj\omega[\Delta_r - (r^2+a^2)\Delta_\theta] + \frac{j^2(a^2\Delta_\theta\sin^2\theta - \Delta_r)}{\sin^2\theta}\right\}$$
$$+\frac{2m\sigma\omega\chi^2}{\Delta_\theta}\left\{(1+\omega)[\Delta_\theta(r^2+a^2)^2 - \Delta_r a^2\sin^2\theta]\right.$$
$$\left.+ja[\Delta_r - (r^2+a^2)\Delta_\theta]\right\} + O(\sigma^2) = 0. \quad (29)$$

Near the event horizon of this black hole, Equation (29) becomes

$$\left(\Delta_r\frac{\partial S}{\partial r}\right)^2\bigg|_{r\to r_H} - \chi^2\left[(r_H^2+a^2)^2\omega^2 + 2aj\omega(r_H^2+a^2) + j^2 a^2\right]$$
$$+ 2m\sigma\omega\chi^2(r_H^2+a^2)[(1+\omega)(r_H^2+a^2) - ja] + O(\sigma^2) = 0, \quad (30)$$

that is

$$\left(\frac{\partial S}{\partial r}\right)^2\bigg|_{r\to r_H} = \frac{\chi^2\left[(r_H^2+a^2)\omega + ja\right]^2}{\Delta_r^2\big|_{r\to r_H}}$$
$$\cdot\left\{1 - \sigma\frac{2m\omega(r_H^2+a^2)[(1+\omega)(r_H^2+a^2) - ja]}{[(r_H^2+a^2)\omega + ja]^2}\right\}$$
$$= \frac{\chi^2 A_0^2}{\Delta_r^2\big|_{r\to r_H}}\left[1 - \sigma\frac{B_0}{A_0^2}\right], \quad (31)$$

therefore,

$$\left.\frac{\partial S}{\partial r}\right|_{r\to r_H} = \pm\frac{\chi A_0}{\Delta_r|_{r\to r_H}}\left[1-\sigma\frac{B_0}{A_0^2}\right]^{1/2}, \quad (32)$$

where,

$$A_0 = (r_H^2+a^2)\omega + ja$$
$$B_0 = 2m\omega(r_H^2+a^2)[(1+\omega)(r_H^2+a^2) - ja]. \quad (33)$$

The positive and negative signs in Equation (32) correspond to the exit wave solution and the incident wave solution respectively. In order to find out the fermion action $S$, we can consider the solution $r_H$ of equation $\Delta_r(r_H) = 0$ as a singularity, so we can integrate at the event horizon of the black hole by applying the residue theorem, and then we can obtain

$$S_\pm = \pm i\pi\left[1-\sigma\frac{B_0}{A_0^2}\right]^{1/2}\frac{\omega - j\Omega_0}{\Delta_r'|_{r\to r_H}(r_H^2+a^2)^{-1}} + S'$$
$$\Omega_0 = -a(r_H^2+a^2)^{-1}. \quad (34)$$

According to the theory of tunneling radiation of the black hole, Equation (34) shows that the quantum tunneling radiation rate of the black hole is

$$\Gamma = \exp\left[-2(ImS_+ - ImS_-)\right] = \exp\left(-\frac{\omega-\omega_0}{T_H}\right), \quad (35)$$

where $T_H$ is modified Hawking temperature at event horizon, expressed as

$$T_H = \frac{r_H - 2r_H^3 l^{-2} - r_H a^2 l^{-2} - M}{2\pi\chi(r_H^2+a^2)}\left[1-\sigma\frac{B_0}{A_0^2}\right]^{-(1/2)}$$
$$= T_0\left[1 + \frac{1}{2}\sigma\frac{B_0}{A_0^2} + \frac{3}{8}\sigma^2\left(\frac{B_0}{A_0^2}\right)^2 + \cdots\right], \quad (36)$$

where $T_0$ is the unmodified Hawking temperature at the event horizon of the black hole. If we ignore the terms after $\sigma$, $T_H = T_0$.

It is worth noting that Equation (36) is an accurate correction based on the Lorenz dispersion relation. This is a modification of the specific theoretical basis, and a small modification on the quantum scale. Since the result of this correction will lead to the correction of black hole entropy, which is related to the information of black hole, the physical significance of this correction is worth studying deeply.

The equation satisfying the cosmic horizon $r_c$ of the black hole is shown in (13). Similarly, by solving the dynamical equation of the arbitrary spin fermions in the space-time of the black hole, we can obtain the fermions tunneling rate at the cosmic horizon $r_c$ of the black hole. Similarly, we can get the Hawking temperature at the cosmic horizon of the black hole, which is expressed as

$$T_H^c = \frac{r_c - 2r_c^3 l^{-2} - r_c a^2 l^{-2} - M}{2\pi\chi(r_c^2+a^2)}\left[1-\sigma\frac{B_0'}{A_0'^2}\right]^{-(1/2)}$$
$$= T_0^C\left[1 + \frac{1}{2}\sigma\frac{B_0'}{A_0'^2} + \frac{3}{8}\sigma^2\left(\frac{B_0'}{A_0'^2}\right)^2 + \cdots\right], \quad (37)$$

where $A_0'$ and $B_0'$ can be got by changing $r_H$ in $A_0$ and $B_0$ into $r_c$ respectively. The difference between getting formula (37) and formula (36) is that when we integrate $\frac{\partial S}{\partial r}\big|_{r\to r_c}$ using the residue theorem, we need to integrate from $r_c$ to the inner where the researcher locates. In other words, $\frac{\partial S}{\partial r}\big|_{r\to r_c}$ and $\frac{\partial S}{\partial r}\big|_{r\to r_H}$ integrate in opposite directions.

Another important physical quantity in black hole thermodynamics is black hole entropy, and the modified Hawking temperature will lead to the modification of black hole entropy. According to the first law of thermodynamics of black holes, the entropy $S^S$ of black holes can be expressed as

$$dS^S = \frac{dM - \Omega dJ - UdQ}{T}. \quad (38)$$

For the Kerr-de Sitter black hole,

$$dS^S = \frac{dM - \Omega dJ}{T}, \quad (39)$$

therefore, through formula (36), the modified entropy at the event horizon of the black hole is expressed as

$$S^S_{r_H} = \int dS^S_{r_H} = \int dS^S_0 \left[1 - \sigma \frac{B_0}{A_0^2}\right]^{(1/2)}$$
$$= S^S_0 \left[1 - \frac{1}{2}\sigma \frac{B_0}{A_0^2} - \frac{1}{8}\sigma^2 \left(\frac{B_0}{A_0^2}\right)^2 + \cdots\right]. \quad (40)$$

Similarly, we can obtain the modified entropy at the cosmic horizon of the black hole, which is expressed as

$$S^S_{r_c} = S^{S'}_0 \left[1 - \frac{1}{2}\sigma \frac{B'_0}{A'^2_0} - \frac{1}{8}\sigma^2 \left(\frac{B'_0}{A'^2_0}\right)^2 + \cdots\right]. \quad (41)$$

In the above two formulas, there are relations $dS^S_0 = (dM - VdJ)/T_0$ and $dS^{S'}_0 = (dM - VdJ)/T^c_0$.

## 4. Conclusion and Discussion

Through a series of calculations, the obtained formulas (35), (36), and (37) show that the tunneling radiation of arbitrary spin fermions in Kerr-de Sitter black hole space-time has been accurately modified based on the Lorenz dispersion theory. The result of this correction shows that when its $a = 0$, the related Hawking temperature at event horizon of black hole returns to the situation of static black hole, and when $\sigma$ and $O(\sigma^2)$ terms are further ignored, the tunneling rate, black hole temperature and other physical quantities return to the situation of Schwarzschild black hole. This further demonstrates the correctness of the conclusions.

The conclusions of Equations (34)–(41) show that Lorenz dispersion relation is a theory worth studying in the field of high energy, and it also should be considered in the study of the theory of strong gravitational field and gravitational wave. The research on these topics can not only promote the research of quantum gravity theory, but also promote the innovation of theoretical physics and astrophysics knowledge.

Moreover, the above calculations show that the tunneling rate $\Gamma$ depends on $T_H$, $\omega$, and $\omega_0 = \Omega_0 j$, but the back reaction from tunneling particles have not be considered. Now let's study the effect. After the black hole emits a particle with energy $\omega$ and angular moment $j$, the mass and angular momentum of the black hole will become $(m - \omega)$ and $(J - j)$, and we have

$$d(M - m) = -d\omega, \quad d(J - j) = -dj, \quad (42)$$

so the tunneling rate is rewritten as

$$\Gamma = \exp\left(-\int \frac{d\omega - \Omega dj}{T_H}\right) = \Gamma$$
$$= \exp\left[\int \frac{d(M-m) - \Omega d(J-j)}{T_H}\right] = \exp(\Delta S). \quad (43)$$

It means tunneling rate is dependent on the entropy of the black hole, so it is a candidate solution of the information loss paradox in black hole physics [45].

On the other hand, we do this work in semiclassical approximation, and in which $\hbar$ is considered as a small number, so that the terms with $\mathcal{O}(\hbar)$ in fold equation is ignored. If we consider the effect from $\mathcal{O}(\hbar)$, the entropy of the black hole will be modified again, and the technology is developed in [15, 16]. We will investigate the effect in future work.

## Data Availability

No data were used to support this study.

## Conflicts of Interest

The authors declare that they have no conflicts of interest regarding the publication of this paper.

## Acknowledgments

We thank Professor Shu-Zheng Yang for his discussion with us on the black hole theory. This work is supported by the National Natural Science Foundation of China (grant number 11273020) and the Natural Science Foundation of Shandong Province, China (grant number ZR2019MA059).